\documentclass[12pt,twoside,reqno]{amsart}
\usepackage{times}
\usepackage{mathptmx}
\usepackage{amssymb}
\begin{document}
\setcounter{page}{1}

\vspace*{1.0cm}

\title[NONLINEAR-DAMPED DUFFING OSCILLATORS HAVING FINITE TIME DYNAMICS] {NONLINEAR-DAMPED DUFFING OSCILLATORS HAVING FINITE TIME DYNAMICS }
\author{  Ronald E. Mickens$^1$, Ray Bullock$^2$, Warren E. Collins$^3$, and Kale Oyedeji$^{4,*}$   }
\date{}
\maketitle

\vspace*{-0.2cm}

\begin{center}
{\footnotesize  $^1$Department of Physics, Clark Atlanta University, Atlanta, GA 30314, USA\\
$^{2,3}$The Center of Physics and Chemistry of Materials, Fisk University, Nashville, TN 37203, USA\\
$^3$Department of Physics, Morehouse College, Atlanta, GA 30314-3773, USA\\

 }
\end{center}
\vskip 2mm

{\footnotesize \noindent {\bf Abstract.}A class of modified Duffing oscillator differential equations, having nonlinear damping forces, are shown to have finite time dynamics, i.e., the solutions oscillate with only a finite number of cycles, and, thereafter, the motion is zero. The relevance of this feature is briefly discussed in relationship to the mathematical modeling, analysis, and estimation of parameters for the vibrations of carbon nano-tubes and graphene sheets, and macroscopic beams and plates.   \vskip 1mm

\noindent {\bf Keywords}: Duffing equation; nonlinear oscillations; finite time dynamics; nonlinear damping.
    2010 AMS Subject Classification: 34C29, 34D05, 70K25, 74H40.

}
  \vskip 6mm

\renewcommand{\thefootnote}{}
\footnotetext{ $^*$'Kale Oyedeji
\par
E-mail addresses:rmickens@cau.edu,rbullock@fisk.edu,ecollins@fisk.edu,and kale.oyedeji@morehouse.edu
\par
Received May 21, 2011

 }

\vskip 6mm
\noindent {\bf\large 1. Introduction}
\vskip 6mm

    The primary purpose of this paper is to demonstrate the  existence of a class of damping forces, such that  when incorporated into the unforced Duffing differential equation \cite{Kovacic1},  the resulting oscillatory motion ends in a finite time. This is in stark contrast with the conclusions reached using damping forces which are linear combinations of terms, each of which are proportional to a positive integer power of the velocity \cite{Kovacic1,Nayfeh1,Mickens1}. For this latter case, in general for large times after the initiation of the oscillation, the amplitude of the oscillation decreases in time either exponentially or by a power law \cite{Nayfeh1, Mickens1}. An example of a very general form for the undamped Duffing equation with a nonlinear damping is given by\cite{Schmidt}

    \vskip 2mm

\begin{equation}
\overset{\cdot \cdot }{x}+\Omega ^{2}x+\varepsilon \beta x^{3}=-\varepsilon %
\left[ d_{0}+d_{1}x^{2}+d_{2}\overset{\cdot }{x}^{2}\right] \overset{\cdot }{%
x}, \tag{1.1}
\end{equation}%

where, for our purposes, all the parameters, $\left( \Omega ^{2},\varepsilon
,\beta ,d_{0},d_{1},d_{2}\right) $ are taken to be non-negative. Note that
the left-side is the standard expression for the Duffing equation, while the
right-side is a damping force which depends on $\ x$ and $\overset{\cdot }{x}%
. $ Also observe that this more extended functional form for the damping force
is even in $x$, but odd in $\overset{\cdot }{x}.$ Thus, in some sufficiently
small neighborhood of the origin, in the (x,y) phase-space, where $y=\overset%
{\cdot }{x}$, the dominant part of the damping force is its linear term,
and, as stated above, the oscillations decrease exponentially. Further,
energy arguments \cite{Mickens2, Jordan} may be used to show that all
solutions of Eq.(1.1), decrease, with increase of time, to zero, i.e.,
\begin{equation}
\underset{t\rightarrow \infty }{\text{Lim}}(x(t),y(t))=(0,0)  \tag{1.2}
\end{equation}

If $0<\varepsilon \ll 1,$ an analytical approximation to the damped
oscillatory solution of Eq.(1.1) may be \ calculated using the method of
first-order averaging \cite{Nayfeh1, Mickens1}. Again, the amplitude has a
monotonic decrease and its magnitude goes to zero only as $t\rightarrow
\infty ,$ in an exponential fashion.

The results to be presented in the remainder of this paper \ are of direct
relevance to the investigation and analysis of several important nonlinear
systems modeled by the Duffing equation. In particular, \ various forms of
this differential equation appear in the study of macroscopic vibrations of
beams and plates \cite{Nayfeh2} as well as the microscopic oscillations of
carbon nanotubes and sheets of graphene \cite{Eichler, Lifshitz, Zaitsev}.

In the next section, a new type of nonlinear damping force is introduced and
its properties are examined. It has the novel feature that it contains a
term for which the velocity is raised to a fractional power, i.e., $f(%
\overset{\cdot }{x})\propto \left\vert \overset{\cdot }{x}\right\vert
^{\alpha },$ where $0<\alpha <1.$ In section 3, a number of general
mathematical comments are discussed. They directly impact the mathematical
computations of this paper and provide clarification as to exactly what is
the proper interpretation which should be given to the existence-uniqueness
theorem in the theory of ordinary differential equations. Section 4 gives
the results of an order $\varepsilon $ analysis of Eq.(2.3) using the method
of averaging. Section 5, discusses possible application of our results and
looks at various limiting cases for the solution to the amplitude equation.
In section 6, general mathematical arguments are presented to show that the
Duffing equation has fininite time dynamics. Finally, section 7 gives a
concise summary and possible \ extension of the work reported in this paper.

\bigskip

\noindent {\bf\large 2. A nonlinear damping force}

\bigskip \noindent Consider a Duffing equation having the following
nonlinear damping force%
\begin{equation}
f(\overset{\cdot }{x})=-\varepsilon \left[ c_{1}\overset{}{\overset{\cdot }{x%
}+c_{2}\text{sgn}(\overset{\cdot }{x}})\left\vert \overset{\cdot }{x}%
\right\vert ^{\alpha }\right] ,  \tag{2.1}
\end{equation}%
where sgn(z) is the "sign" function, i.e.,%
\begin{equation}
\text{sgn}(z)=\left\{ \text{%
\begin{tabular}{ll}
$1,$ & $z>0,$ \\
$-1$ & $z\,<0,$%
\end{tabular}%
}\right.  \tag{2.2}
\end{equation}%
and ($\varepsilon ,c_{1},c_{2})$ are non-negative parameters, with $0<\alpha
<1.$ In normalized form,the corresponding Duffing equation is%
\begin{equation}
\overset{\cdot \cdot }{x}+x+\varepsilon \beta x^{3}=-\varepsilon \left[ c_{1}%
\overset{}{\overset{\cdot }{x}+c_{2}\text{sgn}(\overset{\cdot }{x}}%
)\left\vert \overset{\cdot }{x}\right\vert ^{\alpha }\right] ,  \tag{2.3}
\end{equation}%
where, for our purposes, $\beta $ may be selected to be non-negative.

Given the assumed properties of the various parameters, the function $f(v),$
where $v=\overset{\cdot }{x,}$ has the following features:

(i) $f(v)<0,$ for $v>0;$

(ii) $f(0)=0;$

(iii) $f(-v)=-f(v)$;

(iv) in a neighborhood of the origin, i.e., $v=0,f(v)$ is dominated by the $%
\left\vert v\right\vert ^{\alpha }$ term.

\bigskip

In summary, f(v) is a negative, odd, continuous, nonlinear function, having
the limiting behaviors%
\begin{equation}
f(v)=\left\{ \text{%
\begin{tabular}{ll}
$-\varepsilon c_{1}v,$ & $v$ large$;$ \\
$-\varepsilon c_{2}sgn(v)\left\vert v\right\vert ^{\alpha }$ & $v$\thinspace
small$.$%
\end{tabular}%
}\right.  \tag{2.4}
\end{equation}

\qquad It should be noted that without a fundamental theory to calculate $%
\alpha ,$ the following argument could be used to restrict the
"representation" of $\alpha :$

\qquad (a) Since $f_{\alpha }(v)=-\varepsilon \overset{\cdot }{c_{2}\text{sgn%
}(\overset{\cdot }{x}})\left\vert \overset{\cdot }{x}\right\vert ^{\alpha }$
must be odd and continuous, and further, since it is more "aesthetically
pleasing" not to have the sign-function appear, then what are the
possibilities for $\alpha ,$ if it is required that $0<\alpha <1?$

\qquad (b) Observe that if $\alpha $ is written as
\begin{equation}
\left\{ \text{%
\begin{tabular}{ll}
$\alpha =\frac{2m+1}{2n+1};$ & n, 
m are non-negative integers$;$ \\
$0\leq m\,<n,$ &
\end{tabular}%
}\right.  \tag{2.5}
\end{equation}%
then $0<\alpha <1.$

\qquad (c) Call this set of $\alpha -$values, $\alpha (n,m).$ Then it
follows that
\begin{equation}
\text{sgn}(v)\left\vert v\right\vert ^{\alpha (n,m)}=v^{\alpha (n,m)},
\tag{2.6}
\end{equation}%
and $v^{\alpha (n,m)}$ is, for $v$ real, an odd function of $v$. However,
for arbitrary non-negative real values of $\alpha $, then this term of the
damping force must be written with both the sign-function and the absolute
value operator.

\bigskip

\noindent {\bf\large 3. Mathematical comments}

\bigskip Equation (2.3) can be rewritten as a system of two coupled differential
equations, i.e.,
\begin{equation}
\frac{dx}{dt}=y,\text{ \ \ \ \ }\frac{dy}{dt}=-x-\varepsilon \beta
x^{3}-\varepsilon c_{1}y-\varepsilon c_{2}sgn(y)\left\vert y\right\vert
^{\alpha }  \tag{3.1}
\end{equation}%
In the 2-dimensional (x,y) phase-plane \cite{Mickens1, Jordan, Ross}, this
system has a fixed-point at $(\overline{x},\overline{y})=(0,0),$i.e., the
origin.

A technique to study and analyze the dynamics in the neighborhood of a
fixed-point is to use the method of dominant balance \cite{Bender, Paulsen}.
This general methodology allows the determination of the asymptotic behavior
of the solution by retaining only terms in the relevant equations which are
dominant in the neighborhood of the fixed-point. Since%
\begin{equation}
\left\vert \varepsilon \beta x^{3}\right\vert \ll \left\vert x\right\vert ,%
\text{ \ \ }\left\vert c_{1}y\right\vert \ll \left\vert y\right\vert
^{\alpha },  \tag{3.2}
\end{equation}%
for sufficiently small $x$ and $y$, the dynamics of Eq. (2.3) near the
fixed-point $(\overline{x},\overline{y})=(0,0)$ is determined by the
equations%
\begin{equation}
\frac{dx}{dt}=y\text{, \ }\frac{dy}{dt}=-x-\varepsilon c_{2}sgn(y)\left\vert
y\right\vert ^{\alpha }  \tag{3.3}
\end{equation}%
or%
\begin{equation}
\overset{\cdot \cdot }{x}+x=-\varepsilon c_{2}sgn(\overset{\cdot }{x}%
)\left\vert \overset{\cdot }{x}\right\vert ^{\alpha }.  \tag{3.4}
\end{equation}

\bigskip Another issue to resolve is whether the differential equation%
\begin{equation}
\frac{du}{dt}=-\lambda u^{\alpha },\text{ \ \ }0<\alpha <1,\text{ }\lambda
>0,\text{\ }u(0)>0,  \tag{3.5}
\end{equation}%
has a unique solution. In spite of the fact that the function on the
right-side of Eq. (2.1) does not satisfy the Lipschitz condition \cite{Ross,
Liu}, Eq. (3.5) has a well defined piece-wise, continuous solution given by
the expression%
\begin{equation}
u(t)=\left\{
\begin{tabular}{ll}
$\left[ u_{0}^{(1-\alpha )}-\lambda (1-\alpha )t\right] ^{\left( \frac{1}{%
1-\alpha }\right) },$ & $0<t<t^{\ast },$ \\
$0,$ & $t>t^{\ast },$%
\end{tabular}%
\right.  \tag{3.6}
\end{equation}%
where%
\begin{equation}
u_{0}=u(0)>0,\text{ \ }t^{\ast }=\frac{u_{0}^{(1-\alpha )}}{\lambda
(1-\alpha )}.  \tag{3.7}
\end{equation}%
Examining in detail the standard existence and uniqueness theorem, it states
that if the Lipschitz condition holds, then a unique solution exists.
However, a differential equation not satisfying a Lipschitz condition may or
may not have a unique solution. For the case of Eq. (3.5), a unique solution
exists:

i) For $\lambda >0$ and $u(0)>0,$ the derivative is negative except when $%
u=0.$

ii) $u(t)=0$ is a fixed-point or constant solution of Eq. (3.5).

iii) This result implies that the solutions either decrease, as a function
of $t$, or remain at $u(t)=0$ for $t>0.$

iv) Thus, the unique solution to Eq. (3.5) decreases from $u(0)>0$ to zero,
at $t=t^{\ast }.$ At $t=t^{\ast },$ this solution joins with $%
u(t)=0,$ $ t>t^{\ast }$ to form a continuous solution Equation (3.6), with
condition Eq. (3.7). This is a piecewise continuous solution to Eq. (3.5).

A very concise and clear discussion of these issues is given by Liu \cite%
{Liu}. Also, additional insights on the issue of uniqueness may be gotten
from the book of Kaplan \cite{Kaplan}.

\vskip 6mm

\noindent{ \bf\large  4. Order of $\protect\varepsilon $ and $\protect\varepsilon ^{2}$
analysis of Eq. (2.3)}

\bigskip An approximation to the oscillatory solution of Eq. (2.3) can be determined using the method of first-order averaging (see Mickens \cite{Mickens1}, section3.2). This method takes for the exact solution of the form%
\begin{equation}
x(t,\varepsilon )=a(t,\varepsilon )\cos \left[ t+\phi (t,\varepsilon )\right]
,  \tag{4.1}
\end{equation}%
and under the condition%
\begin{equation}
0<\varepsilon \ll 1,  \tag{4.2}
\end{equation}%
gives the following relations for the first-order differential equations for
the amplitude, $a(t$,$\varepsilon ),$ and the phase, $\psi (t,\varepsilon
)=t+\phi (t,\varepsilon ):$%
\begin{equation}
\frac{da}{dt}=-\left( \frac{\varepsilon }{2\pi }\right)
\int\nolimits_{0}^{2\pi }F(a\cos \psi ,-a\sin \psi )\sin \psi d\psi ,
\tag{4.3a}
\end{equation}%
\begin{equation}
\frac{d\phi }{dt}=-\left( \frac{\varepsilon }{2\pi a}\right)
\int\nolimits_{0}^{2\pi }F(a\cos \psi ,-a\sin \psi )\cos \psi d\psi ,
\tag{4.3b}
\end{equation}%
where for Eq. (2.3)%
\begin{equation}
F(a\cos \psi ,-a\sin \psi )=(ac_{1})\sin \psi +(a^{\alpha })c_{2}\left[
\text{sgn}(\sin \psi )\right] \left\vert \sin \psi \right\vert ^{\alpha
}-\beta a^{3}\left( \cos \psi \right) ^{3},  \tag{4.4}
\end{equation}%
and the initial conditions are
\begin{equation}
a(0,\varepsilon )=A>0,\text{ \ \ }\phi (0,\varepsilon )=0.  \tag{4.5}
\end{equation}%
Note that use has been made of the constraint that the amplitude function
can always be determined in such a way that it is real and non-negative,
i.e.,
\begin{equation}
a(0,\varepsilon )>0\Longrightarrow a(t,\varepsilon )\geqslant 0,\text{ \ }%
t>0.  \tag{4.6}
\end{equation}%
Substitution of Eq. (4,4) into Eq. (4.3), and using the Fourier expansion
relations contained in Kovacic \cite{Kovacic2}, permits the integration of
the expressions appearing on the right-sides of Eq. (4.3):
\begin{equation}
\frac{da}{dt}=-\left( \frac{\varepsilon }{2}\right) \left[
c_{1}a+(b_{1}c_{2})a^{\alpha }\right] ,  \tag{4.7a}
\end{equation}%
\begin{equation}
\frac{d\phi }{dt}=\left( \frac{3\varepsilon \beta }{8}\right) a^{2},
\tag{4.7b}
\end{equation}%
where b$_{1\text{ }}$is \cite{Kovacic2}%
\begin{equation}
b_{1}=\left( \frac{2}{\sqrt{\pi }}\right) \frac{\Gamma \left( \frac{2+\alpha
}{2}\right) }{\Gamma \left( \frac{3+\alpha }{2}\right) }.  \tag{4.8}
\end{equation}

\bigskip

\noindent \qquad Inspection of Eq.(4.7a) shows that it is a Bernoulli
differential equation \cite{Ross} and methods exist to determine its exact
solution \cite{Ross}. If these techniques are applied, then after a
straightforward, but lengthy calculation and further application of careful
mathematical analysis, the following expression is obtained for the
amplitude function%
\begin{equation}
a(t,\varepsilon )=\left\{ \left[ A^{(1-\alpha )}+\frac{b_{1}c_{2}}{c_{1}}%
\right] \exp \left[ -\frac{\varepsilon c_{1}(1-\alpha )t}{2}\right] -\left(
\frac{b_{1}c_{2}}{c_{1}}\right) \right\} ^{\frac{1}{1-\alpha }},  \tag{4.9a}
\end{equation}%
for $\ 0\leq t\leq t^{\ast },$ and

\bigskip

\begin{equation}
a(t,\varepsilon )=0,\text{ \ \ \ }t>t^{\ast },  \tag{4.9b}
\end{equation}%
where%
\begin{equation}
t^{\ast }=\left[ \frac{2}{\varepsilon c_{1}(1-\alpha )}\right] Ln\left[ 1+%
\frac{c_{1}A^{(1-\alpha )}}{b_{1}c_{2}}\right] .  \tag{4.10}
\end{equation}%
Note that $a(t,\varepsilon )$ is a piecewise, continuous, function of t, and
this is true also for its derivative.

The $\phi (t,\varepsilon )$ may be determined by substituting Eqs. (4.9)
into Eq.(4.7b) to obtain
\begin{equation}
\phi (t,\varepsilon )=\left\{
\begin{array}{c}
\left( \frac{3\varepsilon \beta }{8}\right)
\int\nolimits_{0}^{t}a(z,\varepsilon )^{2}\text{d}z,\text{ \ }0\leq t\leq
t^{\ast } \\
\left( \frac{3\varepsilon \beta }{8}\right) \int\nolimits_{0}^{t^{\ast
}}a(z,\varepsilon )^{2}\text{d}z,\text{ \ \ }t>t^{\ast }.%
\end{array}%
\right.  \tag{4.11}
\end{equation}%
However, since the main focus of this paper is on the properties of the
amplitude functions, no calculation of $\phi (t,\varepsilon )$ \ will be
presented. Using the fact that $a(t,\varepsilon )$ is bounded, for all $t,$
it follows from Eqs. (4.11) that
\begin{equation}
\phi (t,\varepsilon )=O(\varepsilon ),\text{ \ \ }t>0.  \tag{4.12}
\end{equation}

Inspection of the argument of the cosine function in Eq. (4.1) and the use
of the result given in Eq. (4.12), implies that the period is
\begin{equation}
T=2\pi +O(\varepsilon ),\text{ \ }0<\varepsilon \ll 1.  \tag{4.13}
\end{equation}%
Therefore an estimate of the number of the oscillations, before the
amplitude becomes zero, is
\begin{equation}
N=\frac{t^{\ast }}{T}.  \tag{4.14}
\end{equation}%
Substituting Eqs. (4.10) and (4.13) into Eq. (4.14) gives
\begin{equation}
N=\left[ \frac{1}{\varepsilon \pi c_{1}(1-\alpha )}\right] Ln\left[ 1+\frac{%
c_{1}A^{(1-\alpha )}}{b_{1}c_{2}}\right]  \tag{4.15}
\end{equation}

Before ending this section, several comments are needed to clarify the
results given in Eqs. (4.9) and (4.10):

\qquad i) \ The first order, nonlinear differential equation, given in Eq.
(4.7a), is a Bernoulli type equation and has the exact solution given by the
expression \ listed in Eq. (4.9a); see Ross \cite{Ross}, pp. 44-46.

\qquad ii) Observe that the right-side of Eq. (4.7a) is negative, since only
$a\geqslant 0$ is physical, and that this function does not satisfy the
Lipschitz condition because of the term $a^{\alpha },$where $0<\alpha <1.$ Never the
less, a solution exists, as given by Eq. (4.9a), and it is unique.

\qquad iii) The uniqueness may be shown by the following argument. Let $%
a(0)=A>0.$ From Eq.(4.7a), it follows that $a(t)$ is monotonic never
increasing function. Further, $a(t)=0$ is a separate (equilibrium) solution.
If $a(t)$ is to be real, then the expression inside the outer brackets, in
Eq. (4.9a), must be non-negative. Since this expression becomes zero at $%
t=t^{\ast },$ where $t^{\ast }$ is given by Eq. (4.10), the solution to our
problem is a piecewise, continuous function, defined by Eqs. (4.9), i.e.,
starting at $t=0,$ with the value $a(0)=A>0,$ the amplitude function
decreases smoothly and monotonically, until at $t=t^{\ast },$ where it
becomes zero; for $t>t^{\ast }$, both $a(t^{\ast })=0$ and $\frac{da(t^{\ast
})}{dt}=0,$ and as a consequence, the complete solution for $a(t)$ is
continuous, as well as its derivative.

\bigskip Finally, we make several comments on the use of the first-order averaging
method as applied to the issues of this paper. The use of dominant balance
reduces Eq. (2.3) to the form given by Eq. (3.4), i.e.,%
\begin{equation}
\overset{\cdot \cdot }{x}+x=-\varepsilon c_{2}sgn(\overset{\cdot }{x}%
)\left\vert \overset{\cdot }{x}\right\vert ^{\alpha }  \tag{4.16}
\end{equation}%
where, as before, $\varepsilon $ and $c_{2}$ are positive parameters, $%
0<\varepsilon \ll 1,$ and $0<\alpha <1.$ The $O(\varepsilon )$ and $%
O(\varepsilon ^{2})$ expressions for the amplitude \ differential equation
are given, respectively, by (see Mickens \cite{Mickens1}, sections 3.4 and
3.5)%
\begin{equation}
\frac{da}{dt}=\varepsilon A_{1}(a),  \tag{4.17a}
\end{equation}%
\begin{equation}
\frac{da}{dt}=\varepsilon A_{1}(a)+\varepsilon ^{2}A_{2}(a).  \tag{4.17b}
\end{equation}%
The function $A_{1}(a)$ was calculated above and shown to be (see the second
term on the right-side of Eq. (4.7a))%
\begin{equation}
A_{1}(a)=-\varepsilon \left( \frac{b_{1}c_{2}}{2}\right) a^{\alpha }.
\tag{4.18}
\end{equation}%
In a similar manner, although the calculations are very intense with
complicated algebraic and trigonometric manipulations, the function $A_{2}(a)
$ can be determined. It turns out that $A_{2}(a)$ is identically zero, i.e.,%
\begin{equation}
A_{2}(a)=0.  \tag{4.19}
\end{equation}%
Thus, both the first- and second-order results for $a(t,\varepsilon )$ are
exactly the same. This means that in a neighborhood of the two-dimensional $%
(x,y)$ phase-space, all the trajectories (of our nonlinear oscillating
system) go to zero to terms of order $\varepsilon ^{2}$, according to the
averaging method.

Note that the time for the oscillations to completely stop is $t^{\ast
}=O(\varepsilon ^{-1}),$ while the validity of the second-order calculations
extend to times $O(\varepsilon ^{-2}).$ This result confirms the conclusion
that the finite-time dynamics is an actual feature of the system.

\bigskip

\noindent {\bf\large  5. General Comments}

\vskip 8mm

It has been shown, for the particular case of the Duffing differential
equation, that $\ $there exists damping forces for which only a finite
number of oscillations take place once the $\ $motions are initiated. While
the damping force, given in this paper, consists of a linear velocity term
and a second one proportional to the velocity raised to a fractional power,
the results are of general applicability. It is to be expected that the
presence of any fractional power term in the damping will lead to
oscillations for which the amplitude goes to zero within a finite time
interval.

In many areas of the natural and engineering services, mathematical models
of the relevant phenomena give rise to Duffing type differential equations.
Any analysis of these models, in conjunction with experimental data, often
allows the estimation of various parameters which are directly related to
physical properties and the dynamics of these systems. However, previous
mathematical models have generally assumed that the damping forces are
either linear or consists of a combination of terms, each proportional to a
positive, odd-integer power of the velocity \cite{Mickens1, Schmidt,
Lifshitz, Zaitsev}. In this case, with no external forcing, oscillations are
found either analytically or by numerical methods to exist for all times
\cite{Nayfeh1, Mickens1, Kovacic2}, i.e., the oscillations take place with
decreasing amplitude for all $t>0$. One, implicit, implication of the
results presented here is that the presence of fractional power damping
terms may yield estimates of system parameter values which differ from the
situation of integer-valued power damping forces. Also this same result
might be expected for systems with external forcing terms.\ For this latter
case, the powerful mathematical techniques created by I. Kovacic \cite%
{Kovacic2, Kovacic3} and A. K. Mallik \cite{Mallik} offer important
possibilities for carrying out the necessary analysis.

It should be noted that one of the important results obtained here is the
determination of an estimate for the finite duration of the oscillations for
the particular damping force given on the right-side of Eq.(2.3). This
estimate is given in Eq.(4.15) and inspection of its mathematical structure
leads to the following conclusions:

i) the number of oscillations depends on all of the parameters appearing in
the damping force; these are$(\varepsilon ,c_{1},c_{2},\alpha )$ and the
initial value of the amplitude, A.

ii) When $\varepsilon $ goes to zero, i.e., there is no damping force, then
the number of oscillations becomes unbounded, and this is exactly what is to
be expected of the undamped harmonic oscillator \cite{Nayfeh1, Mickens1}.

iii) If the limit is taken, where $c_{2}$ goes to zero, then $N\rightarrow
\infty $, a result expected for the case of linear damping. Also, in this
limit, the amplitude function, see, Eq.(4.9a), takes the expected form%
\begin{equation}
a(t,\varepsilon )=Ae^{-\frac{\varepsilon c_{1}t}{2}}.  \tag{5.1}
\end{equation}

iv) If $c_{1}$ is taken to be zero, then using the limit $c_{1}\rightarrow 0$%
, Eqs. (4.9a), (4.10), and (4.15) reduce to the following expressions:%
\begin{equation}
a(t,\varepsilon )=\left\{ A^{(1-\alpha )}-\left[ \frac{\varepsilon
b_{1}c_{2}(1-\alpha )}{2}\right] t\right\} ^{\left( \frac{1}{1-\alpha }%
\right) }\text{ for }0\leq t\leq t^{\ast },  \tag{5.2a}
\end{equation}%
and%
\begin{equation}
a(t,\varepsilon )=0,\text{ \ }t>t^{\ast },  \tag{5.2b}
\end{equation}%
where, for this case%
\begin{equation}
t^{\ast }=\frac{2A^{(1-\alpha )}}{\varepsilon b_{1}c_{2}(1-\alpha )},
\tag{5.3}
\end{equation}%
and, the estimate for the number of oscillations is
\begin{equation}
N=\frac{A^{(1-\alpha )}}{\varepsilon \pi b_{1}c_{2}(1-\alpha )}.  \tag{5.4}
\end{equation}

One point should be clearly understood within the framework of oscillations
as they occur for actual physical systems: In practice, oscillating physical
systems only execute a finite number of oscillations before the amplitudes
of such motions effectively become zero. Thus, a consequence of this fact
should be the imposition on any mathematical model of such phenomena, the
condition that \ the involved forces should be of a nature that the
solutions to the mathematical equations also have this property. The work
presented here strongly indicates that the inclusion in the damping force of
a term having the velocity raised to a fractional power will give this
desired result.

\vskip 6mm
\bigskip
 
\noindent {\bf\large  6. Mathematical consideration of finite time dynamics}

\bigskip The main purpose of the discussion in this section is to provide a more
rigorous demonstration that Eq. (2.3) has finite dynamics. In part, several
of the previous arguments and equations will be listed again.

To begin, consider the Duffing equation with only fractional power damping,
i.e.,
\begin{equation}
\overset{\cdot \cdot }{x}+\varepsilon c_{2}\left[ sgn(\overset{\cdot }{x})%
\right] \left\vert \overset{\cdot }{x}\right\vert ^{\alpha }+x+\varepsilon
\beta x^{3}=0,  \tag{6.1}
\end{equation}%
where, $0<\alpha <1$, $c_{2}>0,$ $\beta \geqslant 0,$ and $\varepsilon >0.$
Note: the parameter $\varepsilon $ is not \textit{a priori} required to be
small; it may take any finite positive value.

This equation has the fixed-point $(\overline{x},\overline{y})=(0,0),$ where
$y=\overset{\cdot }{x}$, and in a sufficiently small neighborhood of this
fixed-point, the method of dominant balance \cite{Bender, Paulsen} allows
consideration only on the following equation%
\begin{equation}
\overset{\cdot \cdot }{x}+\varepsilon c_{2}\left[ sgn(\overset{\cdot }{x})%
\right] \left\vert \overset{\cdot }{x}\right\vert ^{\alpha }+x=0,  \tag{6.2}
\end{equation}%
and this differential equation can be written as a system of two first-order
equations%
\begin{equation}
\overset{\cdot }{x}=y,\text{ \ \ }\overset{\cdot }{y}=-x-\varepsilon c_{2}%
\left[ sgn(\overset{\cdot }{x})\right] \left\vert y\right\vert ^{\alpha }.
\tag{6.3}
\end{equation}

If a polar representation is used for $x(t)$ and $y(t)$, i.e.,%
\begin{equation}
x(t)=r(t)\cos \theta (t),\text{ \ \ \ }y(t)=r(t)\sin \theta (t),  \tag{6.4}
\end{equation}%
then $r(t)$ and $\theta (t)$ satisfy the following first-order differential
equations%
\begin{equation}
\overset{\cdot }{r}=-\varepsilon c_{2}r^{\alpha }\left\vert \sin \theta
\right\vert ^{\alpha +1}  \tag{6.5a}
\end{equation}

\begin{equation}
r\overset{\cdot }{\theta }=-r-\varepsilon c_{2}r^{\alpha }(\cos \theta )%
\left[ sgn(\sin \theta )\right] \left\vert \sin \theta \right\vert ^{\alpha
},  \tag{6.5b}
\end{equation}%
where $r(t)$ and $\theta (t)$, are, respectively, the amplitude and phase
functions. Our interest is primaril
y centered on the amplitude function, $r(t).
$ This function has the following properties:

\bigskip

(i) In the polar representation, $r(t)$ is a non-negative function, i.e., $%
r(t)\geqslant 0.$

(ii) From Eq. (6.5a), it follows that $r(t)$ decreases, since, in general,
its derivative is negative.

\bigskip

(iii) Using energy arguments \cite{Mickens2, Jordan}, it follows that
\begin{equation}
\underset{t\longrightarrow \infty }{Lim}r(t)=0.  \tag{6.6}
\end{equation}%
See Eq. (1.2).

\bigskip

(iv) It is also true, from Eq. (6.5a), that $\overline{r}(t)=0$, is a
nontrivial solution of the nonlinear system of diiferential equations given
by Eqs. (6.5).

\bigskip

Note that independent of the actual time dependence of the phase function, $%
\theta (t),$ the following result holds%
\begin{equation}
0\leq \left\vert \sin \theta \right\vert ^{1+\alpha }\leq 1;  \tag{6.7}
\end{equation}%
are this is the basis for the comment (ii) above. Therefore, the integration
of Eq. (6.5a) gives the result%
\begin{equation}
\left[ r(t)\right] ^{1-\alpha }=(r_{0})^{1-\alpha }-\varepsilon H(t),
\tag{6.8a}
\end{equation}%
where%
\begin{equation}
H(t)\equiv c_{2}\int\limits_{0}^{t}\left\vert \sin \theta (z)\right\vert
^{1+\alpha }dz.  \tag{6.8b}
\end{equation}%
From Eq. (6.7), it follows that $H(t)$ is a montonic increasing function
\cite{Wikipedia} with%
\begin{equation}
H(0)=0,\text{ \ \ }H(t)>0,\text{ \ }t>0,  \tag{6.9}
\end{equation}%
since c$_{2}>0.$ Now there are two possibilities for $H(t)$ \cite{Wikipedia}:

\begin{equation}
\text{(A) \ \ }\underset{t\longrightarrow \infty }{\text{ }Lim}\text{ }%
H(t)=H^{\ast }>0,  \tag{6.10}
\end{equation}

\begin{equation}
\text{(B) \ \ \ \ }\underset{t\longrightarrow \infty }{Lim}\text{ }%
H(t)=\infty .  \tag{6.11}
\end{equation}%
For (A), it follows that
\begin{equation}
\left[ r(\infty )\right] ^{1-\alpha }=(r_{0})^{1-\alpha }-\varepsilon
H^{\ast };\text{ \ \ }r_{0}>0,\text{ given.}  \tag{6.12}
\end{equation}%
But this possibility will, in general not hold. The argument is as follows.
From the energy method, we know that $r(\infty )=0,$ since this is a
dissipative system and the fixed point, $(\overline{x},\overline{y})=(0,0),$
implies that $r(\infty )=0.$ However, regardless of the values $r_{0}$ and $%
H^{\ast },$ a non-zero value of $\varepsilon $ can be chosen such that $%
r(\infty )>0,$ in Eq. (6.12). Thus, as a consequence, possibility (A) can
not hold.

For (B), given any fixed values of $r_{0}>0$ and $\varepsilon >0,$ there
exists a value of $t$, call it $t^{\ast },$ such that

\begin{equation}
(r_{0})^{1-\alpha }=\varepsilon H(t^{\ast }),  \tag{6.13}
\end{equation}%
since $H(t)$ is a monotonic increasing function of $t,$ having the property
stated in Eq. (6.11). Thus, $r(t)$ \ must have the following behavior as a
function of time: At $t=0,$ $r(0)=r_{0}>0.$ From Eqs. (6.5a) and (6.8), $r(t)
$ decreases to zero and achieves the value zero at $\ t=t^{\ast }.$ For $%
t>t^{\ast },$ $r(t)$ is zero. This argument demonstrates that the amplitude
function, $\ r(t),$ is a piece-wise continuous function, with a continuous
first-deruvative.

Observe that the above arguments did not depend on having an explicit
knowledge of $\theta (t)$ as a function of time.

\bigskip

\noindent {\bf\large 7. Summary}

\bigskip We have given arguments in support of the proposition that the use
of nonlinear damping forces, proportional to a fractional power of the
first-derivative, gives rise to dynamics for which the oscillations of
Duffing type equations end in a finite time. This is in contrast to the use
of integer-valued powers of the velocity (first-derivative) which produce
damping dynamics over an arbitrarily large, i.e., infinite time interval.
Using the method of averaging for $\varepsilon $ small, it was shown how to
estimate the magnitude of this time interval. We also provided mathematical
arguments to demonstarte that finite time dynamics are a general feature of
one-dimentional, nonlinear oscillating systems which contain a damping force
having the property that it is \ proportional to the velocity (i.e., first
derivative) raised to a fractional power $\alpha .$

An extension of the work presented here would be to investigate the detailed
properties for the solutions in the case where an external forcing is
included \cite{Kovacic3}. A particular example of such an equation is%
\begin{equation}
\overset{\cdot \cdot }{x}+\Omega ^{2}x+\varepsilon \beta x^{3}=-\varepsilon %
\left[ c_{1}\overset{}{\overset{\cdot }{x}+c_{2}\text{sgn}(\overset{\cdot }{x%
}})\left\vert \overset{\cdot }{x}\right\vert ^{\alpha }+F_{0}\cos (\omega t)%
\right] .  \tag{7.1}
\end{equation}

It should be pointed out that preliminary results on this issue has already
been done in the fundamental paper by Kovacic \cite{Kovacic2}; however, the
possibility of finite time dynamics for the unforced, conservative
oscillations with fractional damping was not examined.

$\qquad $

\bigskip \noindent \noindent
%TCIMACRO{\TeXButton{TeX field}{\noindent}}%
%BeginExpansion
\noindent%
%EndExpansion
\textbf{Acknowledgment\medskip s}

The work of REM was partially supported by funds from the Clark Atlanta
University, School of Arts and Sciences Faculty Development Funds. The work
of RB and WEC was funded by National Science Foundation Grant Award
\#0420516.$\qquad $

\end{document}